\documentclass[journal,comsoc]{IEEEtran}
\IEEEoverridecommandlockouts
\usepackage[T1]{fontenc}
\usepackage[dvips]{psfrag}
\usepackage[dvips]{epsfig}
\usepackage{epic,color}
\usepackage[cmex10]{amsmath}
\usepackage{amsfonts}
\usepackage{amssymb}
\usepackage{amsmath}
\usepackage[cmintegrals]{newtxmath}
\usepackage{bm}
\usepackage{yfonts}
\usepackage{array}
\usepackage{bigstrut}
\usepackage{multirow}
\usepackage{dsfont}
\usepackage{cite}
\usepackage{fixltx2e}
\usepackage{accents}
\usepackage{mathtools}
\usepackage[normalem]{ulem}
\usepackage{enumerate}
\usepackage{stfloats}
\usepackage[T1]{fontenc}
\usepackage{float}
\usepackage{epstopdf}
\usepackage{mathtools}
\usepackage{algorithm}
\usepackage{algorithmicx}
\usepackage{algpseudocode}
\usepackage{algorithm,algpseudocode,float}
\usepackage{lipsum}
\usepackage{soul,xcolor}
\usepackage{xparse}% http://ctan.org/pkg/xparse
\usepackage{acronym}
\usepackage{soul}
\usepackage{graphicx}
\NewDocumentCommand{\ceil}{s O{} m}{%
  \IfBooleanTF{#1} % starred
    {\left\lceil#3\right\rceil} % \ceil*[..]{..}
    {#2\lceil#3#2\rceil} % \ceil[..]{..}
}
\NewDocumentCommand{\floor}{s O{} m}{%
  \IfBooleanTF{#1} % starred
    {\left\lfloor#3\right\rfloor} % \floor*[..]{..}
    {#2\lfloor#3#2\rfloor} % \floor[..]{..}
}

\newcommand{\mbf}[1]{\mathbf{#1}}

\algrenewcommand\algorithmicrequire{\textbf{Input:}}
\algrenewcommand\algorithmicensure{\textbf{Output:}}

\begin{document}

\title{On Performance Comparison of Multi-Antenna HD-NOMA, SCMA and PD-NOMA Schemes}
\author{\noindent \begingroup\centering{Animesh Yadav,~\IEEEmembership{Senior Member,~IEEE}, Chen Quan, Pramod K. Varshney, \IEEEmembership{Life Fellow, IEEE} and H. Vincent Poor, \IEEEmembership{Life Fellow, IEEE}}\endgroup
\thanks{This work was supported in part by the U.S. National Science Foudnation under Grants CCF-0939370 and CCF-1908308. A. Yadav, C. Quan and P. K. Varshney are with Department of Electrical Engineering and Computer Science, Syracuse University, Syracuse, NY, 13244, USA. (e-mail: \{ayadav04, chquan, varshney\}@syr.edu). H. V. Poor is with  the Department of Electrical Engineering, Princeton University, Princeton, NJ 08544, USA.(e-mail: poor@princeton.edu).}}
\maketitle
\begin{abstract}
In this paper, we study the uplink channel throughput performance of a proposed novel multiple-antenna hybrid-domain non-orthogonal multiple access (MA-HD-NOMA) scheme. This scheme combines the conventional sparse code multiple access (SCMA) and power-domain NOMA (PD-NOMA) schemes in order to increase the number of users served as compared to conventional NOMA schemes and uses multiple antennas at the base station. To this end, a joint resource allocation problem for the MA-HD-NOMA scheme is formulated that maximizes the sum rate of the entire system. For a comprehensive comparison, the joint resource allocation problems for the multi-antenna SCMA (MA-SCMA) and multi-antenna PD-NOMA (MA-PD-NOMA) schemes with the same overloading factor are formulated as well. Each of the formulated problems is a mixed-integer non-convex program, and hence, we apply successive convex approximation (SCA)- and reweighted $\ell_1$ minimization-based approaches to obtain rapidly converging solutions. Numerical results reveal that the proposed MA-HD-NOMA scheme has superior performance compared to MA-SCMA and MA-PD-NOMA.
\end{abstract}

\IEEEpeerreviewmaketitle{\noindent }
\vspace{-0.05in}
\section{Introduction}
The non-orthogonal multiple access (NOMA) technique has been identified as one of the spectrally efficient technologies to deal with the current spectrum crisis\cite{fang2016lattice,yadav2019IEEEWCM}. This technique allows multiple users to share the same time-frequency resource for transmission; and consequently, the spectral-efficiency (SE) of the system improves. Several NOMA schemes have been developed, two notable ones being power-domain NOMA (PD-NOMA)\cite{saito2013vtc}, and sparse code multiple access (SCMA) \cite{nikopour2013sparse}. They employ successive interference cancellation (SIC) and the message passing algorithm (MPA), respectively, to remove interference resulting from the other users.

To further improve the SE of the PD-NOMA and SCMA schemes, several works have recently emerged \cite{moltafet2017new,li2016joint,evangelista2019IEEETWC,sharma2019joint}. In \cite{moltafet2017new}, a novel power-domain SCMA (PSMA) scheme for downlink channels is developed by adopting both conventional PD-NOMA and SCMA schemes. This scheme enhances the SE at the cost of increased decoding complexity. In \cite{li2016joint,evangelista2019IEEETWC}, for uplink channels, a joint codebook assignment and power allocation problem for the SCMA scheme to enhance the SE is discussed. In \cite{sharma2019joint}, PD-NOMA and SCMA schemes are combined followed by user grouping for improving the downlink channel capacity. 

Previous works have restrictions on the number of users in the system. However, for the many applications of the Internet of Things (IoT) it is desired to support a large number of users in the system which is also important from the perspective of 5G and beyond networks\cite{chen2020massive}. Further, NOMA when coupled with multiple-antenna technology, which is not considered in the aforementioned works, can enhance the SE of the entire system. In \cite{cquan2020COMML}, the authors have proposed a novel single-antenna hybrid-domain NOMA (HD-NOMA) scheme from the perspective of supporting a large number of uplink users. This scheme combines the SCMA and PD-NOMA schemes where the base station (BS) first categorizes the users into strong user (SU) and weak user (WU) groups based on their channel path losses and then encodes the data of each group according to the SCMA scheme followed by the superposition of the data of the two groups according to the PD-NOMA scheme. At the BS, the SU group is decoded via the MPA and subtracted from the received signal followed by the decoding of the WU group via the MPA. Unlike the single-antenna HD-NOMA scheme in \cite{cquan2020COMML}, in this work we consider a general and complete system model by employing multiple antennas at the BS and joint per-user power and subcarrier allocation to achieve a larger SE. The problem is intractable and challenging in its original from, and hence, a few equivalent transformations and approximations are employed to exploit hidden convexity and improve tractability. The main contributions of this paper are as follows:
\begin{itemize}
    \item A novel multi-antenna HD-NOMA (MA-HD-NOMA) scheme is proposed that supports a larger number of uplink users compared to the previous schemes.
    \item For the MA-HD-NOMA and its comparison with the  multi-antenna SCMA (MA-SCMA) and  multi-antenna PD-NOMA (MA-PD-NOMA) schemes, three joint subcarrier and power allocation design problems are formulated that maximize the sum rate of the entire system.  
    \item Successive convex approximation (SCA)- and reweighted $\ell_1$ minimization- based approaches to handle the non-convex and sparsity constraints, respectively, are used to develop a rapidly converging algorithm.
\end{itemize}

The paper is organized as follows. Section II presents the system model of the conventional and proposed schemes. Section III presents the problem formulations. Section IV discusses the proposed solution and the algorithm. Section V presents and discusses numerical simulation results and Section VI concludes the paper.

\section{System Model}
We consider a single-cell uplink NOMA system where a cell-centered BS equipped with $N_R$ antennas is simultaneously accessed by a set $\mathcal{J}$ of $|\mathcal{J}|=J$ single antenna users using a set $\mathcal{K}$ of $|\mathcal{K}|=K$ orthogonal subcarriers. 
\vspace{-0.05in}
\subsection{Overview of Conventional NOMA Systems}
\subsubsection{MA-CD-NOMA}
In this scheme, a total of $J = \tbinom{K}{d_f}$ users are allowed to access the BS, where $d_f$ denotes the number of subcarriers used by each user out of $K$. Further, at most $d_v=\tbinom{K-1}{d_f-1}$ users are allowed to transmit simultaneously on the same subcarrier. Therefore, user $j$ maps its $\log_2M$ bits onto an $d_f$-sparse $K$-dimensional codeword $\mathbf{x}_{j}=[f_{j,1}s_{j,1}, \ldots, f_{j,K}s_{j,K}]^T$ using its user-specific $M$-elements codebook, where $s_{j,k}$ is the $j$th user symbol transmitted using the $k$th subcarrier, with average power  $p_{j,k}=\mathbb{E}\{|s_{j,k}|^2\}$, and $f_{j,k}=1$ indicates that the user $j$ uses the subcarrier $k$ otherwise it is zero. The complete user-subcarrier association for the SCMA scheme is represented by a factor graph matrix $\mathbf{F}\in \mathbb{B}^{J\times K}$.       

Assuming that the transmission is synchronized, the received signal vector $\mathbf{y}\in \mathbb{C}^{N_RK\times 1}$ at the BS is given as $\mathbf{y} = \sum_{j=1}^{J}\mathbf{H}_j\mathbf{x}_j+\mathbf{n}_{j}$, 
where $\mathbf{n}_j\sim\mathcal{CN}(\mathbf{0}, \sigma_n^2\mathbf{I})$ is the additive white Gaussian noise (AWGN) vector, and $\mathbf{H}_j = [\text{diag}(\mathbf{h}_{j}[1]), \ldots, \text{diag}(\mathbf{h}_{j}[N_R])]^T\in\mathbb{C}^{N_RK\times K}$ is the $j$th user channel matrix with $\mathbf{h}_{j}[n]=[h_{j,1}[n], \ldots, h_{j,K}[n]]^T$ ; $h_{j,k}[n]= g_{j,k}[n]/\text{PL}_j$, where $\text{PL}_j$ is the path loss for user $j$ and $g_{j,k}[n]\sim\mathcal{CN}(0,1)$ denotes the Rayleigh fading channel between the user $j$ and the BS antenna $n$ on the subcarrier $k$. 

To decode the users' data, the BS employs the MPA decoder, which essentially removes the resulting interference. Thus, the achievable sum rate of the MA-SCMA scheme is given as \cite{evangelista2019IEEETWC}
\begin{equation}\label{eq:SR_SCMA}
\text{SR}=\sum_{k=1}^{K}\log_2 \Bigg(1+\frac{\sum_{j=1}^J|\mathbf{h}_{j,k}|^2f_{j,k}p_{j,k}}{\sigma_n^2}\Bigg),
\end{equation} 
where $\mathbf{h}_{j,k} = [h_{j,k}[1],\ldots, h_{j,k}[N_R]]^T$.
\subsubsection{MA-PD-NOMA}
In this scheme, the uplink users with different channel gains superimpose their data and transmit simultaneously to communicate with the BS. In this scheme, for a system with $J$ users and $K$ subcarriers, we assume that each user uses only one subcarrier, i.e., $d_f=1$, and each subcarrier can be accessed by more than one users, i.e., $d_v>=1$. Again assuming synchronized transmission, the received signal vector $\mathbf{y}\in \mathbb{C}^{N_RK\times 1}$ at the BS is given as $\mathbf{y} = \sum_{j=1}^{J}\mathbf{H}_j\mathbf{x}_j+\mathbf{n}_{j}$.
At the BS, the strongest user is decoded first followed by SIC-based decoding to detect the rest of the users that are assumed to be arranged in descending order  of their channel gains.\footnote{ In this work, we assume the following order: J>(J-1)>\ldots >2>1.} The achievable sum rate of the MA-PD-NOMA system under the condition that the SIC decoding is successful is given as \cite{tse2005fundamentals}
\begin{equation}\label{eq:SR_PD_NOMA}
\text{SR}=\sum_{k=1}^{K}\log_2 \Bigg(1+\frac{\sum_{j=1}^J|\mathbf{h}_{j,k}|^2f_{j,k}p_{j,k}}{\sigma_n^2}\Bigg).
\end{equation} 
Although the sum rate expressions for the MA-SCMA and MA-PD-NOMA schemes are the same, the resulting sum rates for the two systems are different as they use different factor graph matrices.  
\vspace{-0in}
\subsection{Proposed Multi-Antenna HD-NOMA Scheme}
\vspace{-0.0in}
In MA-HD-NOMA, the conventional SCMA and PD-NOMA schemes are combined with the aim to support more number of users than supported by the conventional schemes. In order to achieve this, we consider $2J$ users in the set $\mathcal{J}$ and categorized them into two smaller and equal-sized sets  $\mathcal{J}_s$ and $\mathcal{J}_w$ of $|\mathcal{J}_s|=J_s$ SUs and $|\mathcal{J}_w|=J_w$ WUs, respectively, according to the users' large-scale channel fading gains. In this scheme, the users within each group are encoded according to the SCMA scheme whereas the data of the groups are superimposed according to the PD-NOMA scheme because the two groups have different large-scale channel gains.  

Again, assuming synchronized transmission, the received signal vector $\mathbf{y}\in \mathbb{C}^{N_RK\times 1}$ at the BS is given as $\mathbf{y} = \sum_{i=1}^{J_s}\mathbf{H}^s_i\mathbf{x}^s_i+\sum_{j=1}^{J_w}\mathbf{H}^w_j\mathbf{x}^w_j+\mathbf{n}_{j}$, where $\mathbf{H}^s_i = [\text{diag}(\mbf{h}^s_i[1]),\ldots, \text{diag}(\mbf{h}^s_i[N_R])]^T$ and $\mathbf{H}^w_j = [\text{diag}(\mbf{h}^w_j[1]),\ldots, \text{diag}(\mbf{h}^w_j[N_R])]^T$, where $\mbf{h}^s_i$ and $\mbf{h}^w_j$ are defined similarly as $\mbf{h}_j$. At the BS, a two-step decoding process is employed to decode the users' data. In the first step, the MPA is applied on the received signal to decode the users of the SU group while treating the data that belongs to the WU group as noise. In the second step, SIC is performed to remove the data of the SU group from the received signal followed by the MPA on it to decode the users of the WU group.
Thus, the achievable sum rate for the MA-HD-NOMA scheme under the condition that the SU group is decoded correctly is given as \cite{cquan2020COMML}
\vspace{-0.05in}
\begin{IEEEeqnarray}{ll}\label{eq:SR_HD_NOMA}
&\text{SR}_{\text{HD-NOMA}}= \notag\\
&\sum_{k=1}^{K}\log_2 \Bigg(1+\frac{\sum_{i=1}^{J_s}|\mathbf{h}^s_{i,k}|^2f^s_{i,k}p^s_{i,k}+\sum_{j=1}^{J_w}|\mathbf{h}^w_{j,k}|^2f^w_{j,k}p^w_{j,k}}{\sigma_n^2}\Bigg),\quad
\vspace{-0.05in}
\end{IEEEeqnarray} 
where $f^s_{i,k}$ and $f^w_{j,k}$ denote the $(i,k)$th and $(j,k)$th elements of the factor graph matrix corresponding to the SU and WU groups, respectively. $p^s_{i,k}$ and $p^w_{j,k}$ denote the power allocated to the SU $i$ and the WU $j$ on the subcarrier $k$, respectively. We assume that the distances and the channel coefficients from the users to the BS remain constant during the transmission interval.  

Next, we formulate the problem of joint subcarrier and power allocation design so as to maximize the sum rate of the MA-HD-NOMA system. For comparison purposes, we also formulate the joint subcarrier and power allocation problems for the MA-SCMA and MA-PD-NOMA systems with the same number of users as in the MA-HD-NOMA case, i.e, the overloading factors $\delta = 2J/K$ of all the three systems are assumed to be the same. Further, it is assumed that the BS decodes the SU group before the WU group and has full channel state information of all the users to solve the problems in a centralized manner. 

\section{Problem Formulation}
\subsection{MA-HD-NOMA}
The joint subcarrier and power allocation design problem for the MA-HD-NOMA system is written as follow:
%\vspace{-0.5in}
\begin{IEEEeqnarray*}{lcl}\label{eq:HD-NOMA_Orig_P1}
\vspace{-0.1in}
&\underset{\mathbf{F},\mathbf{P}}{\max}\,\, & \text{SR}_{\text{HD-NOMA}} \IEEEyesnumber \IEEEyessubnumber* \label{eq:P1_Obj}\\
&\text{s.t.} & \sum_{k=1}^K \log_2\Bigg(1+\frac{|\mathbf{h}^w_{i,k}|^2f^{w}_{i,k}p^{w}_{i,k}}{\sum_{j=1}^{i-1}|\mathbf{h}^w_{j,k}|^2f^{w}_{j,k}p^{w}_{j,k}+\sigma_n^2}\Bigg) \geq \bar{R}^{w}_i\, \forall i\in \mathcal{J}_w,\notag \\*\label{eq:P1_WU_rate_constr}\\
&& \sum_{k=1}^K\log_2\Bigg(1+\frac{\sum_{i=1}^{J_{s}}|\mathbf{h}^s_{i,k}|^2f^{s}_{i,k}p^{s}_{i,k}}{\sum_{j=1}^{J_{w}}|\mathbf{h}^w_{j,k}|^2f^{w}_{j,k}p^{w}_{j,k}+\sigma_n^2}\Bigg) \geq \bar{R}^{\text{s}}, \label{eq:P1_SU_sumrate_constr}\\
&&\sum\limits_{k=1}^{K}f^{s}_{i,k} \leq d^s_v\quad \forall	i\in \mathcal{J}_s,\quad \sum\limits_{k=1}^{K}f^{w}_{j,k} \leq d^w_v \quad \forall	j\in \mathcal{J}_w  \IEEEyessubnumber*\label{eq:P1_subc_per_UEs}\\
&&\sum\limits_{i=1}^{J_{s}}f^{s}_{i,k} \leq d^s_f\quad \forall k\in \mathcal{K},\quad \sum\limits_{j=1}^{J_{w}}f^{w}_{j,k} \leq d^w_f\quad \forall k\in \mathcal{K} \IEEEyessubnumber* \label{eq:P1_UE_per_subc}\\
&&\sum\limits_{k=1}^{K}f^{s}_{i,k}p^{s}_{i,k} \leq \bar{P}\,\, \forall	i\in \mathcal{J}_s,\quad \sum\limits_{k=1}^{K}f^{w}_{j,k}p^{w}_{i,k} \leq \bar{P} \,\, \forall	j\in \mathcal{J}_w\,\,\, \label{eq:UE_power_constr}\\
&& f^{s}_{i,k}\in \mathbb{B}\, \forall i \in \mathcal{J}_s\,\forall k \in \mathcal{K},\quad f^{w}_{j,k}\in \mathbb{B}\, \forall j \in \mathcal{J}_w\,\forall k \in \mathcal{K},\,\label{eq:binary_constraint}
\end{IEEEeqnarray*}
where $\mathbf{F}$ includes matrices $\{\mathbf{F}^s, \mathbf{F}^w\}$, and $\mathbf{P}$ includes matrices $\{\mathbf{P}^s, \mathbf{P}^w\}$. $\mathbf{F}^s=[[f^s_{1,1},\ldots,f^s_{J_s,1}]^T \cdots [f^s_{1,K},\ldots,f^s_{J_s,K}]^T]$ and $\mathbf{P}^s=[[p^s_{1,1},\ldots,p^s_{J_s,1}]^T \cdots [p^s_{1,K},\ldots,p^s_{J_s,K}]^T]$ and similarly $\mathbf{F}^w$ and $\mathbf{P}^w$ are defined. $\bar{R}^w_i$, $\bar{R}^s$, and $\bar{P}$ denote the minimum rate of the user $i$ in the WU group, the total sum rate of the users in the SU group, and the maximum transmit power of each user, respectively. Constraint \eqref{eq:P1_WU_rate_constr} ensures that the user $i$ in the WU group receives a minimum rate of $\bar{R}^w_i$ whereas the constraint \eqref{eq:P1_SU_sumrate_constr} ensures that the SIC decoding is successful at the BS if the users in the SU group achieve the sum rate of $\bar{R}^s$ b/s/Hz or more. Constraint \eqref{eq:P1_subc_per_UEs} ensures that not more than $d_f$ subcarriers are used by a user in both groups whereas \eqref{eq:P1_UE_per_subc} ensures that not more than $d_v$ users occupy one subcarrier. Constraints \eqref{eq:UE_power_constr} and \eqref{eq:binary_constraint} denote the restrictions on user's maximum transmit power and on each element of the factor graph matrix to be binary for both SU and WU groups, respectively. 
\vspace{-0.15in}
\subsection{MA-SCMA}
In this system, the SU group and the WU group are individually encoded according to the conventional SCMA scheme as previously described and allowed to transmit simultaneously. At the BS, group decoding is performed by treating the other group's data as noise. Hence, the sum rate for this MA-SCMA scheme is given as  
\vspace{-0.1in} 
\begin{IEEEeqnarray}{llc}\label{eq:SR_CD_NOMA}
&\text{SR}_{\text{SCMA}}= &\sum_{k=1}^{K}\Bigg(\log_2 \Bigg(1+\frac{\sum_{j=1}^{J_s}|\mathbf{h}^s_{j,k}|^2f^s_{j,k}p^s_{j,k}}{\sum_{i=1}^{J_w}|\mathbf{h}^w_{i,k}|^2f^w_{i,k}p^w_{i,k}+\sigma_n^2}\Bigg)\notag\\
&& + \log_2 \Bigg(1+\frac{\sum_{i=1}^{J_w}|\mathbf{h}^w_{i,k}|^2f^w_{i,k}p^w_{i,k}}{\sum_{j=1}^{J_s}|\mathbf{h}^s_{j,k}|^2f^s_{j,k}p^s_{j,k}+\sigma_n^2}\Bigg)\Bigg),
\end{IEEEeqnarray} 
and the achievable rate of the user $i$ in the WU group is given as 
\vspace{-0.1in}\small{
\begin{IEEEeqnarray}{ll}\label{eq:UE_rate_CD_NOMA}
\text{R}^w_{i}=\sum_{k=1}^K \notag \\*
\log_2\Bigg(1+\frac{|\mathbf{h}^w_{i,k}|^2f^{w}_{i,k}p^{w}_{i,k}}{\sum\limits_{j=1}^{J_s}|\mathbf{h}^s_{j,k}|^2f^s_{j,k}p^s_{j,k} +\sum\limits_{m=1}^{i-1}|\mathbf{h}^w_{m,k}|^2f^{w}_{m,k}p^{w}_{m,k}+\sigma_n^2}\Bigg).\quad
\vspace{-0.05in}
\end{IEEEeqnarray}}
\normalsize
 
The joint subcarrier and power allocation design problem for the MA-SCMA scheme is written as
\begin{IEEEeqnarray*}{lcl}\label{eq:CD-NOMA_Orig_P1C}
&\underset{\mathbf{F},\mathbf{P}}{\max}\quad & \{\text{SR}_{\text{SCMA}}\mid\text{R}^w_i \geq \bar{R}^{w}_i\, \forall i\in \mathcal{J}_w,\eqref{eq:P1_subc_per_UEs}-\eqref{eq:binary_constraint}\}{\color{blue}.} \IEEEyesnumber\label{eq:P1C_Obj}
\vspace{-0.05in}
\end{IEEEeqnarray*}
\subsection{MA-PD-NOMA}
In this system, the users from the SU and the WU groups are allowed to form a pair to superimpose their signals on a subcarrier. At the BS, the users are arranged in descending order of their channel gains and are decoded using the SIC technique on each subcarrier. Unlike the MA-SCMA, in this system, each user is restricted to use only one subcarrier. Since the BS performs SIC-based decoding, the sum rate expression of the MA-PD-NOMA system is the same as that for MA-HD-NOMA. Therefore, we use \eqref{eq:SR_HD_NOMA} to represent the sum rate of the MA-PD-NOMA system. Furthermore, the achievable rate of the user $i$ in the WU group is given as 
\vspace{-0.05in}
\begin{IEEEeqnarray}{ll}\label{eq:UE_rate_PD_NOMA}
\text{R}^w_{i}=\sum_{k=1}^K 
\log_2\Bigg(1+\frac{|\mathbf{h}^w_{i,k}|^2f^{w}_{i,k}p^{w}_{i,k}}{\sum\limits_{j=1}^{i-1}|\mathbf{h}^w_{j,k}|^2f^{w}_{j,k}p^{w}_{j,k}+\sigma_n^2}\Bigg),
\end{IEEEeqnarray} 
and the achievable sum rate of the SU group is given as
\small{
\begin{IEEEeqnarray}{ll}\label{eq:UE_sumrate_PD_NOMA}
\text{R}^s=\sum_{k=1}^K\sum_{i=1}^{J_s} \notag \\*
\log_2\Bigg(1+\frac{|\mathbf{h}^s_{i,k}|^2f^{s}_{i,k}p^{s}_{i,k}}{\sum\limits_{j=1}^{i-1}|\mathbf{h}^s_{j,k}|^2f^s_{j,k}p^s_{j,k} +\sum\limits_{m=1}^{J_w}|\mathbf{h}^w_{m,k}|^2f^{w}_{m,k}p^{w}_{m,k}+\sigma_n^2}\Bigg).\quad
\end{IEEEeqnarray} }
\normalsize
The joint subcarrier and power allocation design problem for the MA-PD-NOMA system with $d^s_f=d^w_f=1$ and $d^s_v+d^w_v=3$ is written as
\vspace{-0.05in}
\begin{IEEEeqnarray*}{lcl}\label{eq:PD-NOMA_Orig_P1P}
&\underset{\mathbf{F},\mathbf{P}}{\max}\,\, & \{\eqref{eq:SR_HD_NOMA}\mid\text{R}^w_i \geq \bar{R}^{w}_i\, \forall i\in \mathcal{J}_w,\text{R}^s \geq \bar{R}^{s},\eqref{eq:P1_subc_per_UEs}-\eqref{eq:binary_constraint}\}.\quad \IEEEyesnumber\label{eq:P1P_Obj}
\end{IEEEeqnarray*}

Note that, \eqref{eq:HD-NOMA_Orig_P1}, \eqref{eq:CD-NOMA_Orig_P1C}, and \eqref{eq:PD-NOMA_Orig_P1P} are mixed integer non-convex programs due to the presence of non-convex constraints and binary variables and obtaining a rapidly converging optimal solution to each of them is difficult. Hence, we seek suboptimal solutions by developing a rapidly converging algorithm in the following section. The proposed solution method is developed for the MA-HD-NOMA system but is applicable to the other two problems as well.  
\vspace{-0.05in}
\section{Proposed Solution}
In this section, we invoke two steps in an attempt to efficiently solve \eqref{eq:HD-NOMA_Orig_P1}. The first step is about identifying the hidden convexity and achieve tractability by performing some equivalent transformations. Problem \eqref{eq:HD-NOMA_Orig_P1} is non-convex due to the non-linearity of type $xy$ in \eqref{eq:P1_Obj}-\eqref{eq:P1_SU_sumrate_constr} and \eqref{eq:UE_power_constr} and $x/y$ type in \eqref{eq:P1_WU_rate_constr} and \eqref{eq:P1_SU_sumrate_constr}. To tackle these issues, we introduce a set of positive slack variables $\mathcal{V}\in \{z_k, t^s_{i,k},t^w_{j,k},\gamma^{w}_{j,k}, \gamma^s_k,\beta^{w}_{j,k}, \beta^s_k\},\forall i\in \mathcal{J}_s,\forall j\in \mathcal{J}_w,\forall k\in \mathcal{K}$ and rewrite \eqref{eq:HD-NOMA_Orig_P1} equivalently as:
\vspace{-0.05in}
\begin{IEEEeqnarray*}{lcl}\label{eq:HD_NOMA_equi_P2}
&\underset{\mathbf{F},\mathbf{P},\mathcal{V}}{\max}\, & \prod_{k=1}^{K}z_k\quad \IEEEyesnumber \IEEEyessubnumber* \label{eq:P2_Obj}\\
&\text{s.t.} &z_k -1\leq \frac{\sum_{i=1}^{J_s}|\mathbf{h}^s_{i,k}|^2t^s_{i,k}+\sum_{j=1}^{J_w}|\mathbf{h}^w_{j,k}|^2t^w_{j,k}}{\sigma_n^2}, \forall k\in \mathcal{K},\qquad \label{eq:P2_obj_1.1}\\
&& t^s_{i,k} \leq f^s_{i,k}p^s_{i,k} \quad \forall i\in \mathcal{J}_s, \forall k\in \mathcal{K}, \label{eq:P2_nonlin_constr_s}\\
&& t^w_{j,k} \leq f^w_{j,k}p^w_{j,k} \quad \forall j\in \mathcal{J}_w, \forall k\in \mathcal{K}, \label{eq:P2_nonlin_constr_w}\\ 
&&\text{$\prod_{k=1}^K (1+\gamma^w_{j,k})$} \geq 2^{\bar{R}^{w}_j}\quad \forall j\in \mathcal{J}_w,\quad\,\, \label{eq:P2_WU_rate_constr2.1}\\
&& \gamma^w_{j,k}\beta^w_{j,k}\leq |\mathbf{h}^w_{j,k}|^2t^{w}_{j,k} \quad \forall j\in \mathcal{J}_w, \forall k\in \mathcal{K},\label{eq:P2_WU_rate_constr2.2}\\
&& \beta^w_{j,k} \geq \sum\limits_{j=1}^{i-1}|\mathbf{h}^w_{j,k}|^2t^{w}_{j,k}+\sigma_n^2 \quad \forall j\in \mathcal{J}_w, \forall k\in \mathcal{K},\,\label{eq:P2_WU_rate_constr2.3}\\
&& \prod_{k=1}^K (1+\gamma^s_k) \geq 2^{\bar{R}^{\text{s}}}, \IEEEyessubnumber* \label{eq:P2_SU_sumrate_constr3.1}\\
&& \gamma^s_k\beta^s_k \leq \sum\limits_{i=1}^{J_{s}}|\mathbf{h}^s_{i,k}|^2t^{s}_{i,k} \quad \forall k\in \mathcal{K}, \IEEEyessubnumber* \label{eq:P2_SU_sumrate_constr3.2}\\
&& \beta^s_k \geq \sum\limits_{j=1}^{J_{w}}|\mathbf{h}^w_{j,k}|^2t^{w}_{j,k}+\sigma_n^2 \quad \forall k\in \mathcal{K}, \label{eq:P2_SU_sumrate_constr3.3}\\
&&\sum_{k=1}^Kt^s_{i,k}\leq \bar{P}\, \forall i \in \mathcal{J}_s,\quad \sum_{k=1}^Kt^w_{j,k}\leq \bar{P}\quad \forall j \in \mathcal{J}_w,\label{eq:P2_UE_power_constr}\\
&&\eqref{eq:P1_subc_per_UEs},\eqref{eq:P1_UE_per_subc}, \text{ and }\eqref{eq:binary_constraint},\label{eq:P2_all_constr}
\vspace{-0.1in}
\end{IEEEeqnarray*}
where \eqref{eq:P2_Obj} is due to the fact that logarithm is a monotonically non-decreasing function. Problem \eqref{eq:HD_NOMA_equi_P2} is equivalent to \eqref{eq:HD-NOMA_Orig_P1} due to the fact that at the optimal point, all the constraints \eqref{eq:P2_obj_1.1}-\eqref{eq:P2_SU_sumrate_constr3.3} are satisfied as equalities. Hence, the optimal solution to \eqref{eq:HD_NOMA_equi_P2} is also optimal for \eqref{eq:HD-NOMA_Orig_P1}. After the equivalent transformations, \eqref{eq:HD_NOMA_equi_P2} now becomes tractable, however, it is still a non-convex problem due to non-convex and binary constraints present in \eqref{eq:P2_nonlin_constr_s}, \eqref{eq:P2_nonlin_constr_w}, \eqref{eq:P2_WU_rate_constr2.2}, \eqref{eq:P2_SU_sumrate_constr3.2}, and  \eqref{eq:binary_constraint}, respectively. Next, we invoke the second step, which is about relaxing the binary constraints, applying SCA on the non-convex constraints, and developing an iterative algorithm for solving the problem. Inspired by the \textit{reweighted $\ell_1$-minimization} method \cite{candes2008enhancing}, we relax the binary constraint \eqref{eq:binary_constraint} and reformulate the objective function with a penalty term as presented in the following problem at the $n$th iteration:
\begin{IEEEeqnarray*}{lcl}\label{eq:HD_NOMA_equi_P3}
\vspace{-0.1in}
&\underset{\mathbf{F},\mathbf{P},\mathcal{V}}{\max}\, & \eqref{eq:P2_Obj}-\lambda\sum_{k=1}^K\big(\sum_{i=1}^{J_s}w^{s,(n)}_{i,k}f^s_{i,k}+\sum_{j=1}^{J_w}w^{w,(n)}_{j,k}f^w_{j,k}\big)\IEEEyesnumber \IEEEyessubnumber*\label{eq:P3_Obj}\\
&\text{s.t.} &\sum\limits_{k=1}^{K}f^{s}_{i,k} \geq d_v\quad \forall	i\in \mathcal{J}_s,\quad \sum\limits_{k=1}^{K}f^{w}_{j,k} \geq d_v \quad \forall	j\in \mathcal{J}_w \quad \label{eq:P3_subc_per_UEs}\\
&&\sum\limits_{i=1}^{J_{s}}f^{s}_{i,k} \geq d_f\quad \forall k\in \mathcal{K},\quad \sum\limits_{j=1}^{J_{w}}f^{w}_{j,k} \geq d_f\quad \forall k\in \mathcal{K} \label{eq:P3_UE_per_subc}\\
&& 0\leq f^s_{i,k}, f^w_{j,k}\leq 1,\, \forall i\in \mathcal{J}_s,\forall j\in \mathcal{J}_w, \forall k\in \mathcal{K}\label{eq:P3_binary_relax}\\
&& \eqref{eq:P2_nonlin_constr_s}-\eqref{eq:P2_UE_power_constr},\label{eq:P3_all_constr}
\vspace{-0.05in}
\end{IEEEeqnarray*}
where $w^{s,(n)}_{i,k}=1/(|f^{s,(n-1)}_{i,k}|+\epsilon)$ and $w^{w,(n)}_{j,k}=1/(|f^{w,(n-1)}_{j,k}|+\epsilon)$ $\forall i\in \mathcal{J}_s, \forall j\in \mathcal{J}_w, \forall k\in \mathcal{K}$, with a very small value $\epsilon$, and constant $\lambda$ is the weight of the penalty term. Next, to deal with \eqref{eq:P2_nonlin_constr_s}, \eqref{eq:P2_nonlin_constr_w},\eqref{eq:P2_WU_rate_constr2.2} and \eqref{eq:P2_SU_sumrate_constr3.2}, we replace the nonlinear term of type $xy$ by its equivalent difference-of-convex functions, i.e., $xy=0.25((x+y)^2-(x-y)^2)$ followed by linearization of the term $(x+y)^2$ for \eqref{eq:P2_nonlin_constr_s} and \eqref{eq:P2_nonlin_constr_w}, and the term $(x-y)^2$ for \eqref{eq:P2_WU_rate_constr2.2} and \eqref{eq:P2_SU_sumrate_constr3.2} around the point $(x^{(n)},y^{(n)})$ at $n$th iteration. Hence, the final approximate problem we solve at the $n$th iteration is:
\begin{IEEEeqnarray*}{lcl}\label{eq:HD_NOMA_equi_P4}
\vspace{-0.05in}
&\underset{\mathbf{F},\mathbf{P},\mathcal{V}}{\max}\, & \eqref{eq:P3_Obj}\IEEEyesnumber \IEEEyessubnumber*\label{eq:P4_Obj}\\
&\text{s.t.} & t^{s}_{i,k} - \frac{\big(f^{s,(n)}_{i,k}-p^{s,(n)}_{i,k}\big)\big(f^{s}_{i,k} - f^{s,(n)}_{i,k}+ p^{s}_{i,k} - p^{s,(n)}_{i,k}\big)}{2}-\notag\\
&& \frac{\big(f^{s,(n)}_{i,k}+p^{s,(n)}_{i,k}\big)^2}{4} + \frac{\big(f^{s}_{i,k}-p^{s}_{i,k}\big)^2}{4} \leq 0, \, \forall i\in \mathcal{J}_{s}, \forall k\in \mathcal{K},\notag \\*
 \label{P4_non_lin_constr1}\\
&& t^{w}_{i,k} - \frac{\big(f^{w,(n)}_{i,k}-p^{w,(n)}_{i,k}\big)\big(f^{w}_{i,k} - f^{w,(n)}_{i,k}+ p^{w}_{i,k} - p^{w,(n)}_{i,k}\big)}{2}-\notag \\
&&\frac{\big(f^{w,(n)}_{i,k}+p^{w,(n)}_{i,k}\big)^2}{4} + \frac{\big(f^{w}_{i,k}-p^{w}_{i,k}\big)^2}{4} \leq 0, \, \forall i\in \mathcal{J}_{w}, \forall k\in \mathcal{K}, \notag\\*
\label{P4_non_lin_constr2}\\
&& \frac{\big(\gamma^{w}_{i,k} +\beta^{w}_{i,k} \big)^2}{4} - \frac{\big(\gamma^{w,(n)}_{i,k}(n) -\beta^{w,(n)}_{i,k} \big)^2}{4} - |h^{w}_{i,k}|^2t^{w}_{i,k} -\notag \\
&& \frac{\big(\gamma^{w,(n)}_{i,k} -\beta^{w,(n)}_{i,k} \big)\big(\gamma^{w}_{i,k} - \gamma^{w,(n)}_{i,k} -\beta^{w}_{i,k} + \beta^{w,(n)}_{i,k}\big)}{2}\leq 0,\notag\\
&& \qquad\qquad\qquad\qquad\qquad\qquad\qquad \forall i\in \mathcal{J}_{w}, \forall k\in \mathcal{K},\IEEEyessubnumber*\label{P4_non_lin_constr3}\\
&&\frac{\big(\gamma^{s}_{k} +\beta_{k} \big)^2}{4} - \frac{\big(\gamma^{s,(n)}_{k} -\beta_{k}(n) \big)^2}{4} - \sum_{i=1}^{J_{s}}|h^{s}_{i,k}|^2t^{s}_{i,k}- \notag \\*
&&\frac{\big(\gamma^{s,(n)}_{k} -\beta_{k}^{(n)} \big)\big(\gamma^{s}_{k} - \gamma^{s,(n)}_{k} -\beta_{k} + \beta_{k}^{(n)}\big)}{2} \leq 0, \, \forall k\in \mathcal{K},\,\label{P4_non_lin_constr4} \\
&& \eqref{eq:P2_WU_rate_constr2.1},\eqref{eq:P2_WU_rate_constr2.3},\eqref{eq:P2_SU_sumrate_constr3.1},\eqref{eq:P2_SU_sumrate_constr3.3}-\eqref{eq:P2_UE_power_constr},\eqref{eq:P3_subc_per_UEs}-\eqref{eq:P3_binary_relax},\label{eq:P4_all_constr}
\end{IEEEeqnarray*}
Problem \eqref{eq:HD_NOMA_equi_P4} is convex and can be solved iteratively until the value of the objective function converges. The pseudo code that solves the optimization problem is presented in Algorithm~1.
\begin{figure}[h]\label{ALG:SR_SCA}
\vspace{-0.1in}
\centerline{\includegraphics[width=0.9\linewidth,height=3.7 cm]{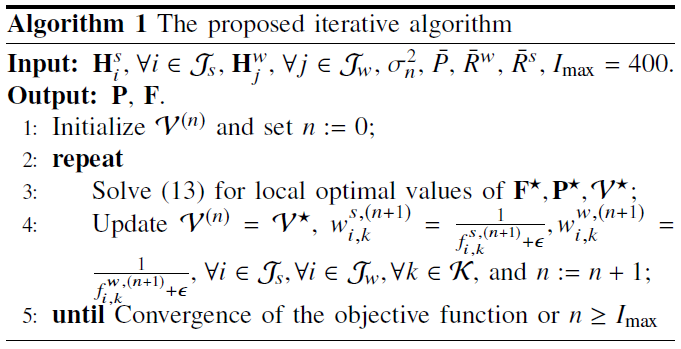}}
\vspace{-0.1in}
\end{figure}
%\small{
%\renewcommand{\baselinestretch}{1}{
%\begin{algorithm}[H]
%\caption{The proposed iterative algorithm}
%\label{ALG:SR_SCA}
%\begin{algorithmic}[1]
%\Require $\mbf{H}^s_i,\, \forall i \in \mathcal{J}_s$, $\mbf{H}^w_j,\, \forall j \in \mathcal{J}_w$, $\sigma_n^2$, $\bar{P}$, $\bar{R}^w$, $\bar{R}^s$, $I_{\text{max}}=400$.
%\Ensure $\mbf{P}$, $\mbf{F}$.
%\State Initialize $\mathcal{V}^{(n)}$ and set $n:=0$;
%\Repeat
%    \State \parbox[t]{\dimexpr\linewidth-\algorithmicindent}{Solve \eqref{eq:HD_NOMA_equi_P4} for local optimal values of $\mbf{F}^{\star}, \mbf{P}^{\star}, \mathcal{V}^{\star}$;}
%    \State \parbox[t]{\dimexpr\linewidth-\algorithmicindent}{Update $\mathcal{V}^{(n)}=\mathcal{V}^{\star}$, $w^{s,(n+1)}_{i,k}=\frac{1}{f^{s,(n+1)}_{i,k}+\epsilon},w^{w,(n+1)}_{i,k}=\frac{1}{f^{w,(n+1)}_{i,k}+\epsilon},\,\forall i\in \mathcal{J}_s,\forall i\in \mathcal{J}_w,\forall k\in \mathcal{K}$, and $n:=n+1$;}
%\Until{Convergence of the objective function or $n\geq I_{\text{max}}$}
%\end{algorithmic}
%\end{algorithm}}} 
%\normalsize
%\section{Convergence and Convexity Analysis}
The objective function in problem \eqref{eq:HD_NOMA_equi_P4} is upper bounded due to the power constraints, and Algorithm~1 generates a monotonic non-decreasing sequence of objective function values, and hence, Algorithm~1 converges to the Karush-Kuhn-Tucker (KKT) point of \eqref{eq:HD_NOMA_equi_P2} \cite{Marks-Wright-or-78}. Since  \eqref{eq:HD_NOMA_equi_P4} admits the second-order conic program (SOCP) due to \eqref{eq:P2_Obj}, \eqref{eq:P2_WU_rate_constr2.1}, and \eqref{eq:P2_SU_sumrate_constr3.1}, Algorithm~1 solves the SOCP in each iteration. The worst case of the complexity is regulated by the SOCP \cite{Lobo-Vandenberghe-Boyd-Lebret-1998} and  is given as $\mathcal{O}(3I(K-1)^3(2+J_w)^2)$, where $I$ is the number of iterations required by the algorithm to converge. 
\begin{figure}[h]
\centerline{\includegraphics[width=0.9\linewidth,height=2.8 cm]{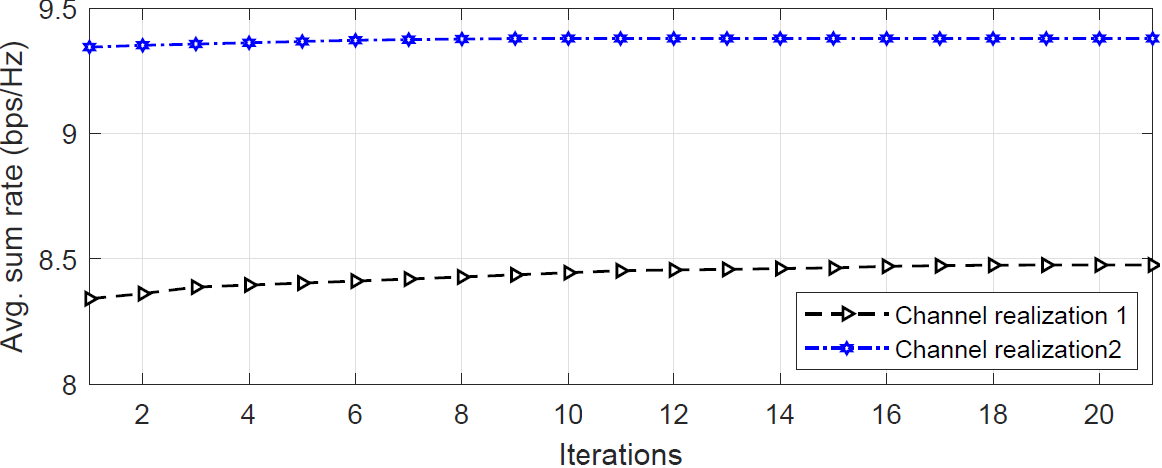}}
\caption{The average sum rate of MA-HD-NOMA versus the number of iterations for  $N_R=4$  and $\bar{P}= 24$ dBm.}\label{fig:convergence}
\vspace{-0.1in}
\end{figure}

\begin{figure}[h]
\centerline{\epsfig{figure=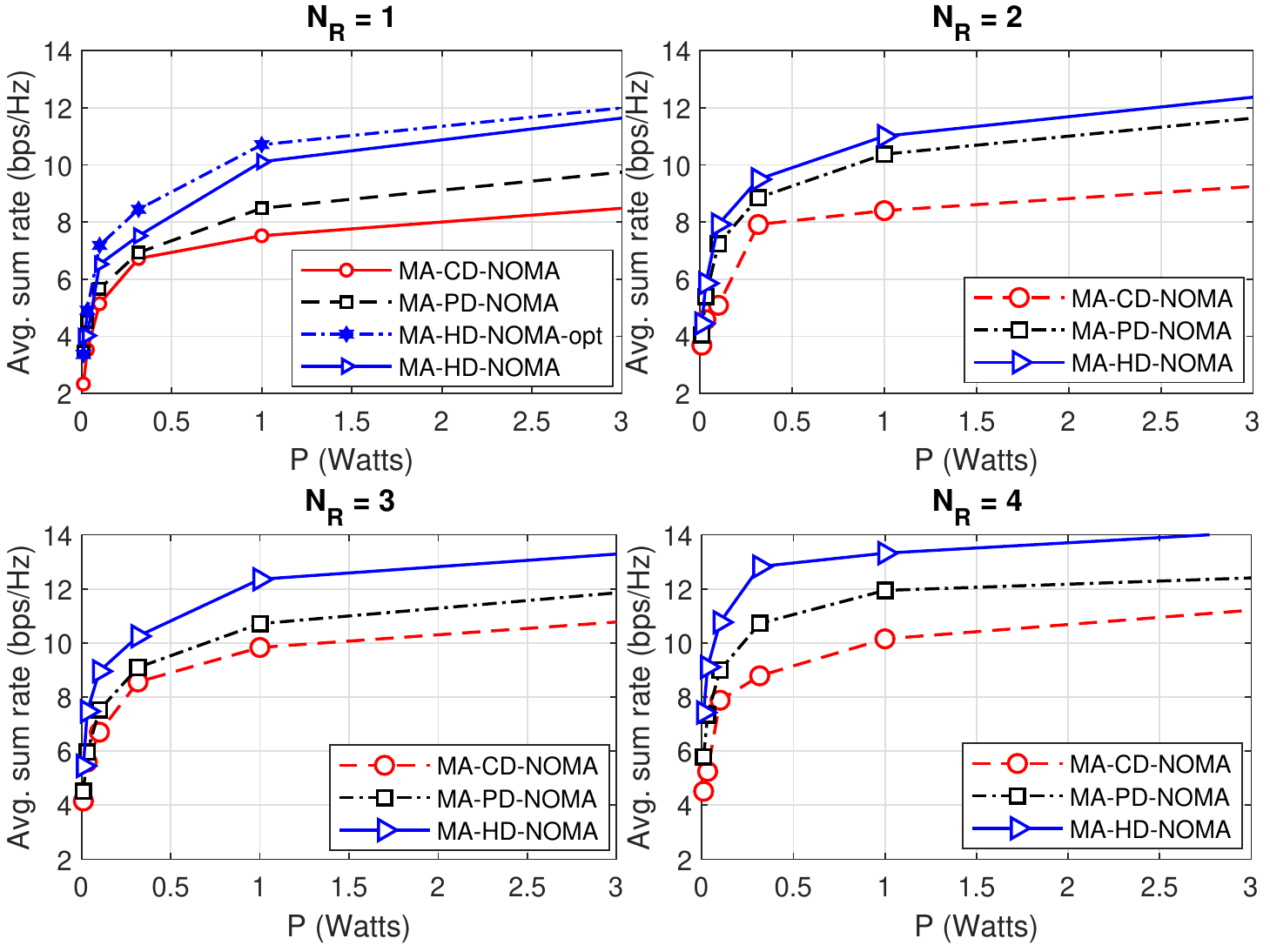,width=2.8in}}
\begin{center}
\caption{The average sum rate versus the maximum transmit power of the users for different number of antennas $N_R=\{1,2,3,4\}$.}\label{fig:SR_vs_Pow}
\end{center}
\vspace{-0.2in}
\end{figure}
\vspace{-0.1in}
\section{Simulation Results}
In this section, we numerically evaluate the sum rate performances of the MA-HD-NOMA, MA-SCMA and MA-PD-NOMA schemes obtained through Algorithm~1 using YALMIP parser and MOSEK as the internal solver.
The users are uniformly distributed within a circular cell of radius $100$ m with the BS at the center. The users that lie within the circular area of radius $d_s\leq50$ m are referred to as the strong users while the others are referred to as the weak users, i.e., $d_w>50$ m. The value of the parameters for the system are set as follows: $\text{BW}=1\,\text{MHz},\,N_R=\{1,2,3,4\},\, K=4,\, J_s=6,\, J_w=6,\, \epsilon=0.001$. Path loss of the user $j$ is denoted as $\text{PL}_j=-145.4+37.5\log_{10}(d)$ dB, where $d$ (in kms.) is the propagation distance of the signal, and noise power is $\sigma_n^2=-204+10\log_{10}(\text{BW}/K)$. $d^s_f=d^w_f=2,\, d^s_f=d^w_f=3$ for the MA-HD-NOMA and MA-SCMA schemes, and $d^s_f=d^w_f=1,\, d^s_f+d^w_f=3$ is set for the MA-PD-NOMA scheme. In Fig.~\ref{fig:convergence}, we plot the average sum rate of the MA-HD-NOMA scheme versus the number of iterations required to obtain a stabilized sum rate for a given random channel realization. It can be observed that Algorithm~1 converges after a few iterations under two different random channel realizations.

%\begin{figure}[h!]
%\centerline{\includegraphics[width=0.9\linewidth,height=4 cm]{figure=Results/convergence.png}
%\begin{center}
%\vspace{-0.1in}
%\caption{The average sumrate of HD-NOMA versus number of iterations for  $N_R=4$  and $\bar{P}= 24$ dBm.}\label{fig:convergence}
%\end{center}
%\end{figure}

In Fig.~\ref{fig:SR_vs_Pow}, we compare the achieved sum rate versus user transmit power of the MA-HD-NOMA scheme to those of the MA-SCMA and MA-PD-NOMA schemes for different values of the number of receiver antennas of the BS. We have three observations from the figure. First, the achieved sum rate of the proposed MA-HD-NOMA system is higher than that of the MA-SCMA and MA-PD-NOMA schemes. Second, the achievable sum rate of all three schemes increases with the number of receiver  antennas. Lastly, the performance of the MA-PD-NOMA scheme is better than the MA-SCMA scheme, which is due to the fact that the MA-SCMA scheme does not perform SIC decoding between the groups at the BS, i.e., the SU and WU group are decoded separately while considering the data from the other group as noise. The optimal solution obtained through a branch-and-bound algorithm is also plotted to compare the performance of the proposed algorithm. For $N_R=1$, we observe that the performance of the MA-HD-NOMA system is quite close to the optimal one.   

\section{Conclusion}
In this letter, we have proposed the MA-HD-NOMA scheme, which has the potential to support more users than MA-SCMA and MA-PD-NOMA, and compared its achievable sum rate performance with these alternatives. The MA-HD-NOMA scheme is a combination of the SCMA and PD-NOMA schemes, and hence benefits from the encoding and MPA-based decoding schemes of the SCMA, and the SIC-based decoding scheme of the PD-NOMA. The subcarrier and power allocation design problem is formulated by maximizing the achievable sum rate for all three systems for a comprehensive comparison. Since the resulting optimization problems are mixed integer non-convex programs, which are difficult to solve, we have proposed a low-complexity approximation of the original problem that rapidly converges to a suboptimal solution. However, the performance of the algorithm is quite close to that of the optimal one. Further, under the same overloading factor, the MA-HD-NOMA scheme outperforms both MA-SCMA and MA-PD-NOMA schemes. Evaluating the performance of the proposed MA-HD-NOMA scheme in multiple cell scenarios is the next logical and viable research direction.
\vspace{-0.2in}
\renewcommand{\baselinestretch}{1}
\bibliographystyle{IEEEtran}

\begin{thebibliography}{10}

\bibitem{fang2016lattice}
D.~Fang \emph{et~al.}, ``Lattice partition multiple access: A new method of
  downlink non-orthogonal multiuser transmissions,'' in \emph{Proc. IEEE Global
  Commun. Conf.}, Washington, DC, USA, Dec. 2016, pp. 1--6.

\bibitem{yadav2019IEEEWCM}
A.~Yadav and O.~A. Dobre, ``All technologies work together for good: A glance
  at future mobile networks,'' \emph{IEEE Wireless Commun. Mag.}, vol.~25,
  no.~4, pp. 10--16, Aug. 2018.

\bibitem{saito2013vtc}
Y.~Saito \emph{et~al.}, ``Non-orthogonal multiple access ({NOMA}) for cellular
  future radio access,'' in \emph{Proc. IEEE Veh. Technol. Conf.}, Dresden,
  Germany, Jun. 2013, pp. 1--5.

\bibitem{nikopour2013sparse}
H.~Nikopour and H.~Baligh, ``Sparse code multiple access,'' in \emph{Proc. IEEE
  Annual Int. Symp. Personal, Indoor, Mobile Radio Commun.}, London, UK, Sept.
  2013, pp. 332--336.

\bibitem{moltafet2017new}
M.~Moltafet \emph{et~al.}, ``A new multiple access technique for 5{G}: Power
  domain sparse code multiple access ({PSMA}),'' \emph{IEEE Access}, vol.~6,
  pp. 747--759, 2017.

\bibitem{li2016joint}
Z.~Li \emph{et~al.}, ``Joint codebook assignment and power allocation for
  {SCMA} based on capacity with {G}aussian input,'' in \emph{Proc. IEEE/CIC
  Int. Conf. Commun. China}, Chengdu, China, Jul. 2016, pp. 1--5.

\bibitem{evangelista2019IEEETWC}
J.~V.~C. Evangelista, Z.~Sattar, G.~Kaddoum, and A.~Chaaban, ``Fairness and
  sum-rate maximization via joint subcarrier and power allocation in uplink
  {SCMA} transmission,'' \emph{IEEE Trans. Wireless Commun.}, vol.~18, no.~12,
  pp. 5855--5867, Jun. 2019.

\bibitem{sharma2019joint}
S.~Sharma, K.~Deka, V.~Bhatia, and A.~Gupta, ``Joint power-domain and
  {SCMA}-based {NOMA} system for downlink in 5{G} and beyond,'' \emph{IEEE
  Commun. Lett.}, vol.~23, no.~4, pp. 971--974, Jun. 2019.

\bibitem{chen2020massive}
X.~Chen \emph{et~al.}, ``Massive access for 5{G} and beyond,'' [Online]
  arXiv:2002.03491v1 [cs.IT], Feb. 2020.

\bibitem{cquan2020COMML}
C.~Quan, A.~Yadav, B.~Geng, P.~K. Varshney, and H.~V. Poor, ``A novel
  spectrally-efficient uplink hybrid-domain {NOMA} system,'' \emph{IEEE Commun.
  Lett.}, vol.~24, no.~11, pp. 2609--2613, Nov. 2020.

\bibitem{tse2005fundamentals}
D.~Tse and P.~Viswanath, \emph{Fundamentals of {W}ireless
  {C}ommunication}.\hskip 1em plus 0.5em minus 0.4em\relax Cambridge University
  Press, 2005.

\bibitem{candes2008enhancing}
E.~J. Candes, M.~B. Wakin, and S.~P. Boyd, ``Enhancing sparsity by reweighted
  $\ell_1$ minimization,'' \emph{Springer J. Fourier {A}nal. {A}ppl.}, vol.~14,
  no. 5-6, pp. 877--905, 2008.

\bibitem{Marks-Wright-or-78}
B.~R. Marks and G.~P. Wright, ``A general inner approximation algorithm for
  nonconvex mathematical programs,'' \emph{Oper. Res.}, vol.~26, no.~4, pp.
  681--683, Aug. 1978.

\bibitem{Lobo-Vandenberghe-Boyd-Lebret-1998}
M.~Lobo \emph{et~al.}, ``Applications of second-order cone programming,''
  \emph{Linear Algebra Appl.}, vol. 284, pp. 193--228, Jan. 1998.

\end{thebibliography}

\end{document}